\documentclass{vgtc}                          % final (conference style)
\ifpdf%                                % if we use pdflatex
  \pdfoutput=1\relax                   % create PDFs from pdfLaTeX
  \usepackage{graphicx}                % allow us to embed graphics files
  \DeclareGraphicsExtensions{.pdf,.png,.jpg,.jpeg} % for pdflatex we expect .pdf, .png, or .jpg files
\else%                                 % else we use pure latex
  \usepackage{graphicx}                % allow us to embed graphics files
  \DeclareGraphicsExtensions{.eps}     % for pure latex we expect eps files
\fi%

\graphicspath{{figures/}{pictures/}{images/}{./}} % where to search for the images

\usepackage{xcolor}
\usepackage{xspace}

\usepackage{microtype}                 % use micro-typography (slightly more compact, better to read)
\PassOptionsToPackage{warn}{textcomp}  % to address font issues with \textrightarrow
\usepackage{textcomp}                  % use better special symbols
\usepackage{mathptmx}                  % use matching math font
\usepackage{times}                     % we use Times as the main font
\usepackage{cite}                      % needed to automatically sort the references
\usepackage{tabu}                      % only used for the table example
\usepackage{booktabs}                  % only used for the table example
\onlineid{0}

\vgtccategory{Research}

\vgtcinsertpkg

\title{Goals, Process, and Challenges of \\\ Exploratory Data Analysis: An Interview Study}

 \author{Kanit Wongsuphasawat\thanks{e-mail: kanitw@apple.com. This work was done when the first author was at the University of Washington.}\\ %
         \scriptsize Apple Inc. %
 \and Yang Liu\thanks{e-mail: yliu0@cs.washington.edu}\\ %
      \scriptsize University of Washington %
 \and Jeffrey Heer\thanks{e-mail: jheer@uw.edu}\\ %
      \parbox{1.4in}{\scriptsize \centering University of Washington}}

\newcommand{\verified}[1]{#1}

\newcommand{\yang}[1]{}

\newcommand{\LATER}[1]{}

\newcommand{\ie}{{\em i.e.}\xspace}
\newcommand{\eg}{{\em e.g.,}\xspace}

\newcommand{\etal}{{\em et~al.}\xspace}

\newcommand{\commonvspace}{\vspace{4pt}}

\newcommand{\paperonly}[1]{#1}
\newcommand{\thesisonly}[1]{}
\newcommand{\paper}{paper\xspace}
\newcommand{\tabemph}[1]{\emph{#1}}
\newcommand{\secref}[1]{\hyperref[#1]{\S\ref*{#1}}}
\newcommand{\figref}[1]{\hyperref[#1]{Fig.~\ref*{#1}}}

\renewenvironment{quote}{%
   \list{}{%
     \leftmargin0.45cm   % this is the adjusting screw
     \rightmargin\leftmargin
   }
   \item\relax
}
{\endlist}

\newcommand{\q}[1]{
\vspace{-5.5pt}
\begin{quote}
\emph{``#1''}
\end{quote}
\vspace{-5.5pt}
}

\newcommand{\inlineq}[1]{\emph{``#1''}}

\abstract{

How do analysis goals and context affect exploratory data analysis (EDA)?
To investigate this question,
we conducted semi-structured interviews with 18 data analysts.
We characterize common exploration goals:
\emph{profiling} (assessing data quality) and \emph{discovery} (gaining new insights).
Though the EDA literature primarily emphasizes discovery,
we observe that discovery only reliably occurs in the context of open-ended analyses, whereas all participants engage in profiling across all of their analyses.
We describe the process and challenges of EDA highlighted by our interviews.
We find that analysts must perform repetitive tasks (\eg examine numerous variables),
yet they may have limited time or lack domain knowledge to explore data.
Analysts also often have to consult other stakeholders and oscillate between exploration and other tasks, such as acquiring and wrangling additional data.
Based on these observations, we identify design opportunities for exploratory analysis tools, such as augmenting exploration with automation and guidance.
}

\CCScatlist{
  \CCScatTwelve{Human-centered computing}{Visu\-al\-iza\-tion}{Visu\-al\-iza\-tion techniques}{Treemaps};
  \CCScatTwelve{Human-centered computing}{Visu\-al\-iza\-tion}{Visualization design and evaluation methods}{}
}

\begin{document}


\firstsection{Introduction}

\maketitle

%% \section{Introduction} %for journal use above \firstsection{..} instead
% !TEX root =  eda-interview.tex

\newcommand{\figureCoding}{
  \begin{figure*}[t!]
  \centering
  \vspace{-10pt}
  \includegraphics[width=.82\textwidth]{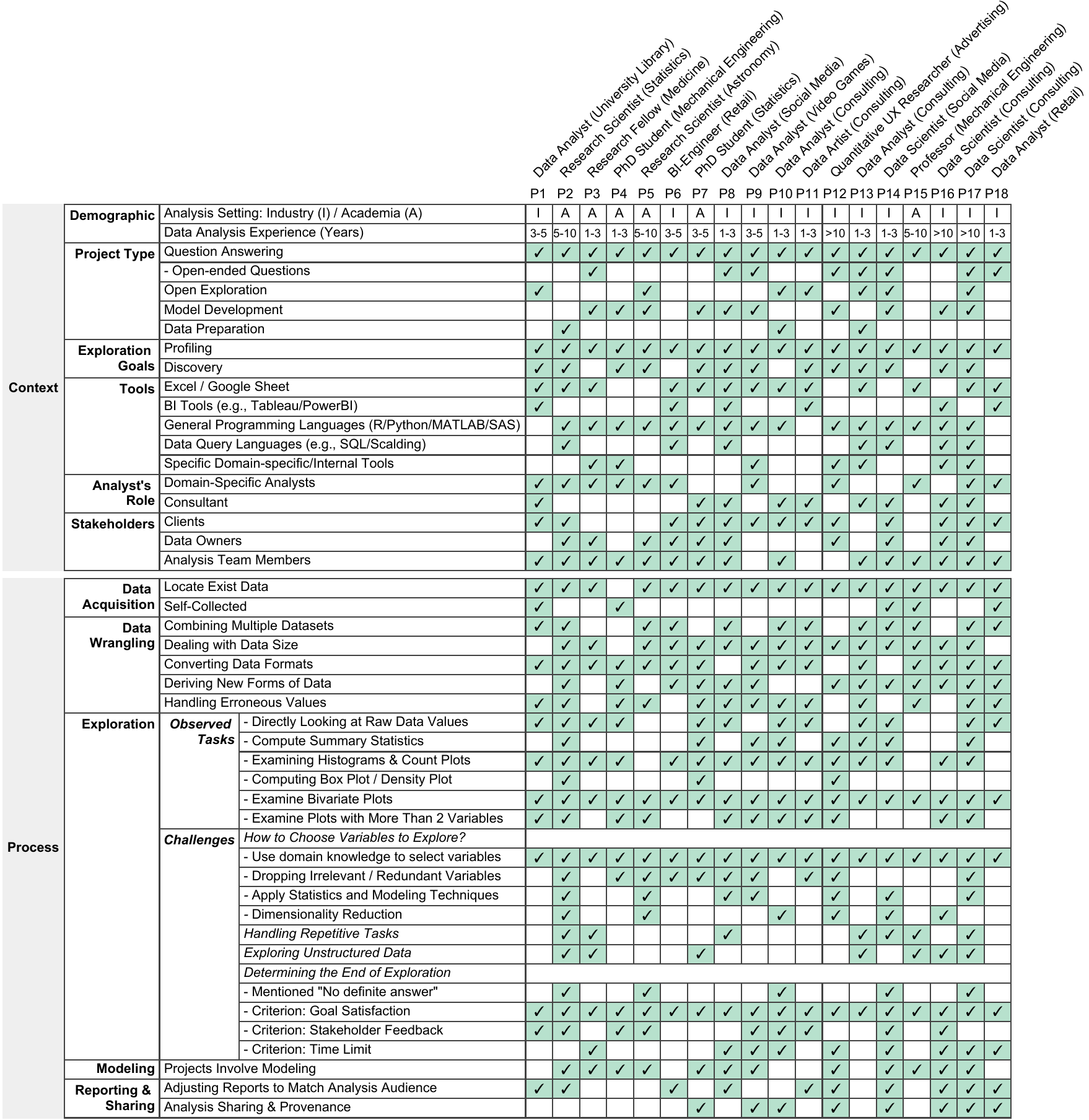}
  \vspace{-3pt}
  \caption{\label{fig:coding}
    A matrix of interviewees, their corresponding analysis context, and high-level tasks they perform in the analysis process.
  }
  \vspace{-8pt}
  \end{figure*}
}

\newcommand{\figureProcess}{
  \begin{figure*}[t!]
  \vspace{-5pt}
  \centering
  \includegraphics[width=.85\textwidth]{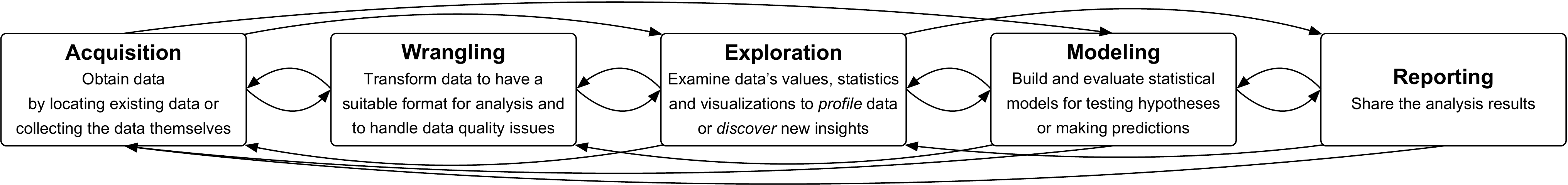}
  \vspace{-7.5pt}
  \caption{\label{fig:process}
    The analysis process couples exploration with many tasks including acquisition, wrangling, modeling, and reporting.
  }
  \vspace{-10pt}
  \end{figure*}
}

% !TEX root =  eda-interview.tex

Exploratory data analysis (EDA), as introduced by Tukey~\cite{tukey:eda},
aims to complement formal confirmatory analysis with
a ``flexible attitude'', letting data exposure inform analysts' modeling decisions~\cite{tukey:weneedboth}.
With this attitude, analysts usually ``explore'' % different
aspects of data by examining data values, derived statistics, and visualizations.
% While Tukey's original practice mainly relied on manual calculation and hand-drawn graphics,
% many % in Statistics, Human Computer Interaction, and Information Visualization.
% have since developed a number of exploration software tools and techniques (\eg
% \cite{wickham:ggplot,shneiderman:eyes,polaris,card:infovis,cleveland:graphing-data,few:nowyouseeit}).
Today, data exploration is widely adopted as a critical part of data science,
both in industrial and scientific settings~\cite{idreos:exploration}.
% Though analysts perform data exploration in various kinds of analyses,
% little research has observed how analysis goals and context affect
% the day-to-day practice and challenges of exploratory analysis.
However, while analysts perform data exploration in various kinds of analyses,
the EDA literature lacks a consistent definition of exploration goals.
Moreover, little research has observed how analysis goals and context affect
the day-to-day practice and challenges of EDA.
Understanding these issues can inform the design of % improved
data exploration tools.

To better understand current EDA practices,
we conducted semi-structured interviews with 18 analysts from academic
and industrial settings.
We asked the analysts to describe their analysis goals, tasks they performed,
and challenges they faced in their exploration.
% This paper presents the results and analysis of these interviews.
We first describe observed analysis context and process.
% From the interviews, we find that analysts often couple exploration with other tasks including acquisition, wrangling, and reporting.
We discuss observed types of analyses that involve exploration
and identify two common exploration goals: % the characteratization of two common exploration goals:
\emph{profiling} (understanding  what the data contain and assessing data quality)
and \emph{discovery} (gaining new insights).
Though the EDA literature emphasizes discovery, we observe that
all participants engage in profiling across all of their analyses,
while discovery only reliably occurs in open-ended analyses, which participants perform less often.
Based on the participants' descriptions of their analysis process,
we revise Kandel et al.'s model of the data analysis process~\cite{kandel:enterprise} to include exploration.
We also report the analysts' context including tools, domain knowledge (or the lack thereof), and involved stakeholders.

Next, we discuss recurring observed challenges in the data analysis process
and report how analysis goals and context impact them.
We also describe how analysts handle challenges specific to exploration tasks % (and their interdependencies)
including choosing variables to explore, handling repetitive tasks, and determining the end of an exploration.
We find that analysts often have to explore numerous variable combinations, requiring them to apply domain knowledge to select and reduce the number of variables.
As analysts perform repetitive tasks, they may curate analysis templates
to automate their routines and help them follow best practices.
% While exploring data, analysts often have to switch to other tasks
% including data acquisition and wrangling, as well as consulting with and reporting to other stakeholders.
Due to time limits, analysts may also need to move on to
other tasks before completing their exploration.

Finally, we identify opportunities for data exploration tools.
We argue that tools can help mitigate these observed challenges and
facilitate rapid and systematic exploration by providing automation for
routine tasks and guiding analysis practices.
We also note a lack of support for data wrangling
and navigation of analysis history within exploration tools.

% Our contributions include:
% (1) the characterization of exploration goals (profiling and discovery),
% (2) details of how analysis goals and context affect EDA tasks, and
% (3) identification of challenges specific to exploration tasks including how analysts choose variables to explore, handle repetitive tasks, and determine to end an exploration.

% !TEX root =  eda-interview.tex

\section{Background and Related Work}

We build on the exploratory data analysis literature and
complement prior work on understanding data analysis.

\subsection{Exploratory Data Analysis}

Exploratory data analysis stems from the collection of work by the statistician John Tukey
in the 1960s and 1970s~\cite{jones:tukey3,jones:tukey4,cleveland:tukey5,tukey:eda}.
%Many of these seminal articles were compiled in
%the collection of Tukey's works, especially in Volume 3-5.
His seminal book~\cite{tukey:eda} compiles a collection of data visualization techniques
as well as robust and non-parametric statistics for data exploration.
Many communities including Statistics, Human-Computer Interaction, and
Information Visualization have since contributed new data exploration tools
and techniques
(\eg \cite{wickham:ggplot,shneiderman:eyes,polaris,card:infovis,cleveland:graphing-data,few:nowyouseeit}).
\thesisonly{(See Chapter 3 for detailed reviews of related data exploration tools.)}

\newcommand{\edaisdiscovery}{\cite{behrens:eda,behrens-yu:eda,filliben:eda,gelman:complexeda,morgenthaler:eda,leinhardt:eda,seltman:experimental,hoaglin:robusteda,velleman:abc,andrienko:eda-spatial-temporal}\xspace}

While Tukey did not explicitly define the goals of EDA, his and other
writings about EDA~\edaisdiscovery mostly focus on the discovery of structure and
patterns in the data, and consider EDA a step that precedes formal modeling or
confirmatory analysis. However, some \cite{chatfield:ida,chatfield:eda,wickham:r}
argue that EDA also covers profiling~\cite{abedjan:profiling,naumann:profiling},
or initial data examination to detect data quality issues.
Some also state that EDA may occur without formal modeling~\cite{chatfield:problemsolving}.
As the prior literature lacks a consistent definition of EDA goals,
our study helps clarify the nature and scope of EDA
by providing evidence that EDA goals include both profiling and discovery.
Though the EDA literature emphasizes discovery,
we observe that discovery only reliably occurs in open-ended analyses, while
all participants engage in profiling across all analyses.
We also find that some analysts perform exploration to clean or summarize
data without modeling involved.
Besides characterizing goals, we also identify common challenges for
data exploration and discuss how analysis goals and context affect them.
% We complement the EDA literature by reporting
% observed EDA challenges and discussing how analysis goals and context affect them.

\subsection{Understanding Data Analysis}
\label{sec:interview-rel}

Many prior studies summarize high-level tasks and challenges in the data analysis process.
Some focus on specific user groups and types of analysis.
Kwon and Fisher~\cite{chul:va-novice} discuss visual analytic challenges for novices.
Conversely, we study experts whose jobs primarily involve data analysis.
A few studies~\cite{chin:intelligence,kang:characterizing,pirolli:sensemaking}
examine data analysis and sensemaking within intelligence agencies, which share many challenges
with our findings due to exploratory and collaborative nature of their work.
However, these agencies often analyze text documents whereas
our participants mostly explore structured data.
For structured data, Guo~\cite{guo:thesis} describes research programming practices while Fayyad \etal~\cite{fayyad:knowledge} discuss the process of algorithmic data mining.

Another group of works~\cite{kandel:enterprise, wickham:r, alspaugh:futzing} discusses general data analysis process.
% However, Wickham and Grolemund~\cite{wickham:r} do not provide
% empirical evidence to support the described process.
Closest to our work are the interview studies by Kandel \etal~\cite{kandel:enterprise} and Alspaugh~\etal~\cite{alspaugh:futzing},
which derive analysis tasks and challenges based on interview data.
However, Kandel \etal mostly focus on analysts that perform directed analyses
(\ie answering predefined questions) and
largely overlook tasks and challenges specific to discovery. Meanwhile, Alspaugh \etal
interview analysts about exploratory activities akin to this work,
but focus only on open-ended exploration and do not discuss how analysis goals affect exploration challenges.
In contrast, this study covers exploratory activities across open-ended and directed analyses.
We complement these prior studies with the characterization of exploration goals (profiling and discovery) and details of how analysis goals and context affect EDA tasks
and challenges.
In~\secref{sec:process}, we also discuss how our characterization of the data analysis process
differs from those of Kandel \etal and Alspaugh \etal

Some prior studies investigate specific issues in data analysis, \eg
the effects of latency~\cite{liu:latency} and
multiple comparisons~\cite{zgraggen:mcp}.  Some study specific tools such as
computational notebooks~\cite{rule:notebook,kery:notebook},
interactive visualizations~\cite{batch:interactive-gap,yi:interaction},
and dashboards~\cite{sarikaya:dashboard}.
In contrast, we study day-to-day practices of EDA,
which involve many challenges and tools.
For exploration challenges,
Lam~\cite{lam:interaction-cost} discusses interaction costs for visualizing data
such as repetitive physical motions and choosing data subsets.
% such as choosing data subsets and performing repetitive physical motions.
Kidd~\cite{kidd:marks} observes that knowledge workers often focus on implications
for decision-making rather than producing generalizable knowledge.
Others examine low-level tasks for visual exploration.
Amar \etal~\cite{amar:task} present a taxonomy of low-level
visual analytics tasks. A few studies~\cite{jankun:model,chi:operator}
% Jankun-Kelly \etal~\cite{jankun:model} and Chi and Riedl~\cite{chi:operator}
identify operations that analysts perform to visualize data.
Conversely, we focus on the high-level data exploration process.

Prior work also discusses some of the analysis challenges observed in our study.
Many studies (\eg~\cite{elmagarmid:dedup,kandel:directions,kandel:enterprise,rahm:datacleaning,keim:challenges})
discuss challenges for data wrangling such as data integration,
data cleaning, and handling large data.
Here we discuss how data wrangling couples with and impedes exploration.
% 12, 15 in Sean's, Burrito
% Some \TODO{Fink} discuss that domain specific tools often have limited interoperability with other tools.
% \TODO{Some~\cite{keim:challenges} note the difficulties the need for domain knowledge.}

Many researchers~\cite{callahan:vistrails,gotz:characterizing,jankun:model,kandel:enterprise,ragan:provenance} have also noted the importance of analytic provenance.
Ragan \etal~\cite{ragan:provenance} also characterize types and purposes of provenance in visual analytics. Some studies~\cite{isenberg:exploratory,kang:characterizing,kandel:enterprise}
identify how and why analysts collaborate, and discuss impediments for collaboration.
Rule \etal~\cite{rule:notebook} also describe the tension between exploring
data and documenting insights for computational notebook users.
Batch \etal~\cite{batch:interactive-gap} also comment that
visualization tools lack integration with data science workflows.

% \TODO{label intel?}
% \TODO{Petra~\cite{isenberg:exploratory}}
% \TODO{Reda \etal~\cite{reda:modeling}}
% \TODO{insight based, sense.us}

Though this study shares some findings with prior research,
our work is the first, to our knowledge, to overview the day-to-day process
and challenges of EDA for both profiling and discovery aspects,
including examination of how analysts choose variables to explore and
determine when an exploration should stop.

\section{Methods}

To better understand day-to-day practices of exploratory data analysis, we conducted semi-structured interviews with experienced analysts across both academia and industry.

\figureCoding

\subsection{Participants}

%
%We interviewed 18 analysts (11 male, 7 female) from both academia and industry.
%Six participants came from two universities, working in fields including
%Astronomy, Oceanography, Medicine, Statistics, and Mechanical Engineering.
%Another participant was on a data analysis team of a university library.
%The remaining interviewees were from seven software companies,
%working on a variety of topics including video-gaming, safety and trust,
%real estate logistics, and advertising.
%The participants' job titles included "UX researchers", "Data Artist", "Data Scientists",
%"Data Analysts", "BI-Engineer", "Professor", "Research Fellow", "Research Scientist", and "Ph.D. Student".

We interviewed 18 analysts (11 male, 7 female) from both academia and industry.
As listed in~\figref{fig:coding}, the participants worked on various research fields
and industrial topics, and held a variety of job titles.
% Six of them included a professsor, researchers, and PhD students from two universities and five fields including
% Astronomy, Oceanography, Medicine, Statistics, and Mechanical Engineering.
% One participant was a data analyst for a university library.
% The remaining participants were industry analysts from seven software companies,
% holding job titles including UX Researcher, Data Artist, Data Scientist,
% Data Analyst, and Business Intelligence Engineer.
% These industrial analysts worked on various topics including video-gaming,
% social media, retail, and advertising.
In this \paper, we use the term ``analyst'' to generally refer to any participant,
as all participants' jobs primarily involved data analysis.

To recruit participants, we emailed our contacts within
our personal and professional networks to forward our recruiting emails to analysts
in their organizations. We used a survey to screen participants to those that
had at least one year of data analysis experience and performed EDA
at least once a month.
The participants' data analysis experience varied from 1-3 years
to over 10 years. Most of them performed EDA
on a daily or weekly basis, with the least frequent account being biweekly.
While our recruitment strategy introduced potential sampling bias in the results,
our primary goal is to characterize the space of day-to-day
exploratory analysis process and challenges, not to quantify how frequently
each specific task occurs. To better quantify these results, other methods,
such as surveys, could complement our findings.

\subsection{Interview}

We conducted semi-structured interviews with one interviewee at a time.
Each interview lasted from 45 to 90 minutes. We interviewed analysts
at their workplace when possible, and used video calls otherwise.
For each interview, we began by describing the study objective,
namely to understand current practices and difficulties of exploratory data analysis.
We then asked open-ended questions and encouraged interviewees to describe
their specific experiences such as ``walk us through a recent exploratory data analysis scenario.''
Our questions aimed to learn about the following topics:

\vspace{-5pt}

\begin{itemize}
\setlength\itemsep{-0.3em}
\item What are your analysis goals and outcomes?
% Where do their data typically come from and what are their typical data type and size?
\item What tasks do you perform during analyses?
\item What tools do you use and how do you use them?
\item How do you interact with other involved stakeholders?
% \item What are the kinds of data sources and formats?
\item How do you choose parts of a dataset to explore?
\item How long does an exploration take?
\item How do you decide that an exploration is complete?
% How confident are they that they have comprehensively explore/analyze the data?
\item What are the key challenges you face in exploratory analysis and how do you handle them?
% \item What are things they wish for in data exploration?
\end{itemize}

\subsection{Analysis}

We analyzed the interview data using an iterative coding method.
\paperonly{The first two authors}\thesisonly{Yang Liu and I} independently coded all data.
Throughout the coding process, we discussed disagreements and iteratively
revised our codes to ensure consistency across coding sessions.
% From the data, we categorized common practices
% and challenges in the exploratory analysis process, as well as the strategies
% to handle these challenges.
The rest of this \paper presents the results from this analysis (summarized in \figref{fig:coding}).
We also include representative quotes from the interviews to support these results.
We use P1-P18 to refer to the participants.

% !TEX root =  eda-interview.tex

\section{Analysis Process and Context}
\label{sec:process}

From the interview responses, we first categorize analysis projects based on
their overarching objectives and identify two kinds of exploration goals.
We then report observed high-level tasks in the analysis process.
We also discuss analysts' context including tools, their operational
and domain knowledge, and their collaboration with involved stakeholders.

\subsection{Types of Analysis Projects}

We asked the interviewees about the objectives % and outcomes
of the projects that involved exploratory analysis.
We observed four common project types, with varying levels of open-endedness.

\commonvspace

\tabemph{Question Answering}.
All analysts (18/18) reported working on answering business and research questions,
so they explored the data to check data quality before answering them.
Many analysts (8/18) also noted that their questions, while predetermined,
were sometimes open-ended and thus required exploration to discover answers,
as P14 said:

\q{A lot of my work is more long-term open-ended research questions such as: how can we characterize the health of the users on our platform?}

% \q{A lot of my work is more long term open-ended research questions such as: how can we characterize the health of users who are using the platform?}

Analysts often produced analysis reports in the form of written documents and
presentation slides. They also sometimes built interactive dashboards.

\commonvspace

\tabemph{Open-Ended Exploration}. While answering specific questions was more common,
several analysts (7/18) noted that they sometimes broadly explored data to summarize and
look for new insights \emph{without} a specific question.
P17, a data science consultant, reported that his clients once gave him
their website's data and asked \inlineq{Please just tell me about my site.}
P5, an astronomer, also said:

\q{Occasionally we get data that's surprising like the universe does something we haven't seen before and a telescope caught it. Then you sit down with the data and think `What do I do now?'}

Akin to question answering, analysts often produced reports to describe
insights from the open-exploration process.

% P17 You may go into this without a clear goal in mind. You may not even know what it is you're looking for. Just tell me please about my site. Where are we gaining customers, losing customers? We have a weird robot here. There are just stuff. There’s all sorts of information there. What parts of our site are inefficiently designed and we should shorten it or we should improve the content in this part of the site? You may have some goal in mind where it’s like, oh, we ran some promotion. How effective was that promotion? How did that affect customer behavior? You may have some particular study objective in mind for your analysis, but it’s still a very free-form semi-structured data analysis problem.

\commonvspace

\tabemph{Model Development}. Many analysts (10/18) reported cases
where they performed exploratory analysis to prepare for modeling projects
such as training machine learning models or developing new metrics and rules.
Besides the models, analysts might also deliver reports, or integrate the solutions into dashboards as their project outcomes.

\commonvspace

\tabemph{Data Publishing}. A few analysts (\verified{3}/18) explored data while
cleaning datasets for publishing on shared repositories, so others could
use the datasets for other analyses.
% Besides the published data, two analysts
% often included computational notebooks to demonstrate example analyses on the data.

\figureProcess

\subsection{Exploration Goals}

We asked the analysts why they performed data exploration in their analysis projects.
From their descriptions, we categorize their goals into two common categories:

\commonvspace

\tabemph{Profiling}. A common goal for all analysts (18/18) was to learn
what the data contained and assess if the data were suitable for the analyses.
By broadly looking at the data and their plots, analysts could
learn about their shapes and distributions, and detect data quality issues
such as missing data, extreme values, or inconsistent data types.
They might also check specific assumptions of the data, both
in terms of expectations based on domain knowledge and
mathematical assumptions required for modeling.
By profiling, they learned if the data were ready for the analyses
or if they needed to further wrangle the data or acquire more data.

%We observed that analysts verified two types of data assumptions.
%First, they verified whether the data shape and distribution
%matched their \emph{assumptions based on prior domain knowledge}.
%For example, one analyst anticipated that product sales would increase during
%Thanksgivings and Christmas. Another expected the count of social media followers
%by users to have zero-inflated log-normal distributions.
%Second, they checked if the data satisfied \emph{mathematical assumptions} required for
%statistical models they planned to use. For example, one analyst checked if the data
%items are independent.  Another check if the data points in a time series were evenly
%sampled. Some assumption checks also triggered analysts to further wrangle the data.
%For example, a few analysts reported applying log transformation to make the derived
%data more uniformly distributed.

%\q{Any time working with data, the hardest thing is if it's inconsistent or it's missing. It always goes back to, wherever you got the data from, is it reliable? Can you find who is the owner of the data or who reported it? Just the basic stuff that anyone would go with data, which is saying, can you trust it? Is it reliable? Is it biased? Those types of things, and making sure that when you put it in there, it's the best data that you can get.}

%As one said:
%\q{Exploratory analysis is just to get a hunch and just to see that your dataset is fine. There are some signals which are not ridiculously distributed, just to make sure that everything is fine or not, and to get an idea of how the signals are distributed}

\commonvspace

\tabemph{Discovery}. Many analysts (13/18) also explored data to discover
new insights or hypotheses,
%such as potential relationships between variables,
as P17 described that his exploration goal was to
\inlineq{be open-minded and learn what the data could tell me.}
% \inlineq{be as open-minded as I could and learn what the data could tell me.}
For question answering and modeling projects, analysts might focus on
developing intuitions how to answer questions or formulate models such as
learning about potential relationships between variables or rankings of feature importance.
Some insights also inspired the analysts to broadly explore other relevant factors
while some helped them form and investigate specific questions.

% \TODO{P17: For production / streaming data, learning different ways that
% data can have error -- so they can provide safeguard}

%\inlineq{What is in the data? I try to have as open a mind as I can. What can the data tell me?}

\commonvspace

Analysts' focus on exploration goals depended on project objectives.
While the EDA literature (reviewed in \secref{sec:interview-rel})
% (\eg~\cite{behrens:eda,morgenthaler:eda,hoaglin:robusteda,tukey:eda,velleman:abc})
mostly focuses on discovery,
we observed that profiling was a more common goal.
Projects with fixed questions generally centered on profiling,
though surprising observations from profiling sometimes prompted
analysts to investigate and discover the causes of the surprises.
Meanwhile, open-ended analyses involved both goals.
Analysts often first focused on profiling, and shifted their focus to
discover new insights when they felt more confident about the data.

% Ham: Wonder if we need to explicitly say that there is no clear
% boundary between these two goals and they may still interleave.
% (Perhaps not, since we already says "shift the focus", which means that
% is not the only thing they do)

%\q{“For example if it's financial data for how much stuff the store has sold, I got the idea that the sales should have a weekly pattern, or a peak during that time of the year that is usually associated to higher [inaudible 00:09:20] like Christmas time or Thanksgiving time.”}

\subsection{High-Level Tasks in the Analysis Process}

From the interviewees' responses about the tasks they performed in their analyses,
we characterize the data analysis process as an iterative process that couples
five common high-level tasks: \emph{acquisition}, \emph{wrangling},
\emph{exploration}, \emph{modeling}, and \emph{reporting}
(as listed and defined in~\figref{fig:process}).
%
% \commonvspace
%
% \tabemph{Acquisition}---obtaining data either by locating existing data
% or collecting the data themselves.
%
% \tabemph{Wrangling}---transforming data to have a suitable format for analysis
% and to handle data quality issues.
%
% \tabemph{Exploration}---examining data by looking
% at the data values as well as their metadata, statistics, and visualizations.
%
% \tabemph{Modeling}---building and evaluating statistical models for testing hypotheses or making predictions.
%
% \tabemph{Reporting}---sharing the analysis results.
%
% \commonvspace
%
Some projects might omit some tasks.
For example, though exploration often preceded modeling, some analysts (6/18)
explored data to clean or summarize data without modeling involved.
Some data were also clean and did not require wrangling.

The analysts' process coupled exploration with many tasks.
The analysts regularly explored data to assess if the data were
relevant during acquisition. Similarly, they often explored data
to decide how to wrangle them. Exploration also helped them
discover the need to collect or wrangle more data.
In addition, the analysts often reported exploration results to
other stakeholders and gathered feedback for more exploration.
While we observed less coupling between modeling and exploration, a few
analysts examined training data when they observed poor modeling results.

% (akin to ~\cite{kandel:enterprise,wickham:r}):
Our characterization of analysis tasks is similar to those of Kandel \etal~\cite{kandel:enterprise}
and Alspaugh \etal~\cite{alspaugh:futzing}.
% Following Kandel \etal, we categorize the analysts' activities as high-level tasks.
However, as Kandel \etal focus on analysts that typically perform directed analyses
(answering predetermined questions), they only list \emph{profiling}
rather than \emph{exploration} as one of the tasks.
Alspaugh \etal, whose study focuses on open-ended analyses,
augment Kandel \etal's model by adding \emph{exploration} as an alternative task to \emph{modeling}.
In contrast, as our study covers exploratory tasks for both directed and open-ended analyses,
we found that analysts often explored data prior to modeling.
They also often performed similar exploration tasks
(examining the data's values and derived statistics and visualizations, as described in \secref{sec:ex-tasks})
to profile data or discover new insights.
Thus, we revise Kandel \etal's model by replacing \emph{profiling} with a more general \emph{exploration} task, which subsumes both profiling and discovery goals.

In \S\ref{sec:acquisition}-\ref{sec:reporting}, we discuss
common challenges in these tasks and report how analysts handled them.
Though analysts also explored variations of models and outputs,
this \paper focuses on data exploration. We consider model diagnostics beyond
the scope of this \paper.

\subsection{Analysis Tools}
\label{sec:tools}

The interviewees reported using and switching between multiple tools throughout their analyses.
A few (P1, P11, P18) were application users who usually looked at and wrangled data
in spreadsheets, and visually explored data in Tableau. % P1, P11, P18
% P1 also used OpenRefine to wrangle data.

The rest (15/18) were programmers who primarily used one language among
Python, MATLAB, R, and SAS to analyze data. They usually plotted data with APIs
such as Matplotlib~\cite{matplotlib} and ggplot2~\cite{wickham:ggplot}.
Several of them also used computational notebooks (\eg Jupyter~\cite{perez:jupyter})
to keep history for repeating and revising their analyses.
Some noted that they preferred exploring data via scripting instead of using
graphical interfaces as they did not have to switch tools.
However, the programmers switched to other tools in some cases.
P6 sometimes explored data in Tableau when it could connect to the data sources.
Several used spreadsheets to inspect raw data, though they rarely wrangled data
in spreadsheets like the application users.
Many utilized languages such as SQL and Scalding to fetch and manipulate the data.
Some used Tableau~\cite{polaris}, Google Data Studio~\cite{googledatastudio},
or Microsoft PowerBI~\cite{powerbi} for reporting.

% To share their analysis data, code, and results, the analysts used
% collaboration tools including emails, instant messaging software,
% shared drives (\eg Google Drive), version control software (\eg Git),
% and internal wikis.

% Analysts cited several criteria for choosing tools. Several Python and R users
% praised the availability of analysis libraries in their ecosystems.
% Meanwhile, several analysts chose tools used by their communities or their teams.
% P14 noted that she preferred R but used Python to facilitate code sharing as her teammates used it, while  P2 used R since it was the standard for her community.
% Another researcher (P2) mentioned that she used R because it is familiar
% for her community, though it was not the most efficient tool for her.
% \q{People are familiar with it. Even though it's not the most efficient way to do things all the time, it's more readable and understandable in the field.}

Several analysts sometimes had to use domain-specific tools.
% A few researchers reported using tools specific to their research fields.
P3 explored biopsy images from a 3D scan with
a specialized tool. A few industrial analysts also noted that
their internal data platforms had some support for data wrangling and exploration.
As domain-specific tools often had limited features,
analysts preferred to use general-purpose tools if possible.
However, their data often resided in domain-specific tools
and exporting data was sometimes difficult.

\subsection{Operational and Domain Knowledge}

The analysts typically needed operational and domain knowledge in their analyses.
They must know where the data were stored, and
how the data were collected and processed. They also needed domain
expertise to interpret the data and detect errors.
Since analysts usually lacked some required knowledge,
they had to learn more about the problem domains and consult other stakeholders.

%\q{I have to do things like look up literature on whatever I have data of to see if there is something that can guide me or talk to collaborators about it}

Job roles also affected the levels of operational and domain knowledge that analysts had.
We observed that the analysts had two kinds of job roles relative
to their problem domains: \textit{domain-specific analysts} (9/18)
and \textit{consultants} (7/18), with two (2/18)
straddling both roles in different phases of their careers.
In academia, most researchers focused on their research topics, but one
(P7) was a statistician providing solutions to multiple research domains.
In industry, there were both analysts embedded into product teams and
consultants who served internal or external clients.
As consultants typically worked with a broader set of domains,
they often had less domain expertise and relied more on
other stakeholders, as P17 said:

%\q{Because I'm not embedded with the team, I don't necessarily have all the domain context. In this example where I saw elevated counts in the product's telemetry,
%I didn't know what that meant. I could make guesses, but I'm not on the team, so I have no idea.}{P17}

\q{Since I'm not embedded with the team, I don't have the domain context.
In this example where I saw elevated counts in the product's telemetry,
I didn't know what it meant. I could guess, but I'm not on the team, so I have no idea.}

\subsection{Stakeholders and Collaboration}

We observed that analysts collaborated with a few types of stakeholders
over the course of their analysis projects.

\commonvspace

\tabemph{Clients}.
Most analysts (13/18) had clients who prompted
them to perform the analyses and were the direct audience for the results.
Some analysts were consultants who served external clients while
some worked with internal clients within their organizations such as
product managers or executives.
Analysts often interacted with clients in an iterative fashion.
Besides reporting the final results, analysts might share preliminary results
and ask the clients for feedback such as verifying if the results matched
the clients' prior knowledge and checking if the analyses aligned with the project goals.

% the clients' prior knowledge, checking if the analyses aligned with the project goals,
% and asking if the clients had any additional questions.

\commonvspace

\tabemph{Data owners}.
Many analysts (10/18) interacted with data engineers or
database administrators who curated, processed, or stored the data prior to
their analyses. Clients were also sometimes data owners, directly providing
the data for the analysts. Analysts often asked the data owners to provide
additional information to help them locate, clean, and understand the data
since data owners had a better understanding of the format and meaning
of the data as well as where the data were stored and how they were processed.

%\ham{There is also data engineers in consulting team (who help analysts fetch data / do SQL / etc.) -- I guess we can ignore that.}

\commonvspace

\tabemph{Analysis Team Members}.  % Fellow analysts?
Though the analysts primarily analyzed data on their own,
most of them (15/18) were members of analysis teams. Thus, they regularly obtained
feedback from fellow analysts and supervisors %via team meetings and informal conversations
before presenting to clients.
%\ham{Twitter person: obtained feedback within the team first before presenting to clients}
Typical feedback included additional questions to explore, technical advice
for analysis techniques and implementation, and suggestions
to make the reports easier to understand for the clients.
Moreover, a few interviewees noted that they worked jointly with
their colleagues on some projects. Two reported splitting the work so
each team member could focus on an independent scope and make progress in parallel.
Another mentioned that she and her colleague independently analyzed
the same data and cross-checked if they arrived at the same results.

Besides supervisors and fellow analysts, a few interviewees had colleagues
with more domain expertise in their teams. P3's medical device research team
had a pathologist to give opinions on tumor image analysis.
P16, a data science consultant, also reported that his organization
included business-oriented ``solution managers'', whose duties were to
bridge the communication gap between the clients and technical-oriented
data scientists and help them define deliverables that matched the clients' goals.

% \TODO{Let's categorize this as a project manager}
% As clients often lacked expertise in data science while some data scientists
% were sometimes not familiar with business domains, one data scientist
% reported that his consulting team had a role called ``solution managers''
% for business-oriented team members that served as intermediaries
% between data scientists and their clients.

% P16 -- one of the responsibilities of the solution manager is to help define those deliverables and help making sure that they are delivered at the right time and right format.

%  \q{they need to be up to date about what we are going to present to the client. And they help with filling the gap that sometimes the client is not very savvy in data science, so they help ... they are more savvy of the business, then the solution manager brings in the language that they understand better and they help communicating between the data scientists and the client.}

%\TODO{What are the things that they share? (Kandel says data, scripts, results, and documentation)}

% !TEX root =  eda-interview.tex

\section{Data Acquisition}
\label{sec:acquisition}

We now discuss challenges for data exploration and relevant activities.
The first step is to acquire the data necessary for the analysis.
%, either by using existing datasets or by collecting new data.
\verified{All but one} interviewee (17/18) reported working with
existing datasets. For business analysts, most data were from product logs
or customer surveys, while many researchers worked on datasets
jointly collected by their research communities. Only some (5/18)
had participated in data collection, either by collecting
the data themselves or requesting that certain data should be collected.

% Ham: There is one case where analyst augment existing data too,
% but I think that's not important so let's drop it.

When working with existing data, finding relevant data
were difficult for a few reasons. First, data were often distributed.
Several analysts reported that their companies used multiple data storage
infrastructures. A few researchers also mentioned that their datasets
were collected and published by different research organizations.
Thus, analysts typically had to search for data in many places.
Moreover, data sources often had insufficient data description,
having uninformative column labels and missing or outdated documentation.
As a result, analysts had to explore all potential
datasets to assess if they were relevant to their analyses.

%One analyst also commented that he preferred that the data pipeline
%be clean and ready to use without the need to read documentation
%but he rarely got datasets with such property in practice.

%One analyst also expressed his frustration about documentation:
%\q{I generally prefer to have the data pipeline be clean and clear to
%use. If you need a document to explain how to use it, you've failed.}{}
%\TODO{Read the quote and expand}

Some analysts % (5/18) % Some industry analysts
consulted data owners to locate and understand the data.
They often received connections to the data owners from their clients or colleagues.
However, P14 noted that finding the right people to talk to was difficult
since she worked in a remote office.

% P14 -  A lot of the difficulties I face are actually more to do with the fact
% that I work in our little office. Finding good sources of data is always like
% a challenge because we have to work like closely with engineers as
% instrument. Their products or so on and sometimes they haven’t and it’s like
% sometimes difficult to find like the right person to talk to.

% Analysts considered establishing
% relationship with their data owners is an important part of their jobs.
% However, this can be difficult for analysts who worked remotely as
% it was harder for them to establish relationship with the right group of people.

%\q{The first part of this process is to ask around people and be like, ``Okay you work on the [team name]''. Is there some place easily accessible where this data is located?}

Analysts also used keyword search to look for relevant datasets in their databases.
However, as the same data could be named in many ways, they had to try
many different keywords to find the data.
For some analysts, their data sources might not have convenient search
capability at all. Due to this problem, P2 noted that she
was building a searchable database for her organization.

For consulting analysts, they often received their data from the clients. However,
the provided data might lack the information necessary to achieve
the project goals, requiring the analysts to search for more appropriate data or otherwise terminate the project.
% if they could not find appropriate data.

% P1 - Yes, there are times where I do exploratory analysis where I’m not confident in the results. I am usually a lot more skeptical of the data than the tools though. I think a lot of that is because we have people doing ad hoc projects and they’re just send us their stuff and it doesn’t seem like either the project design is matched up with the kinds of questions they’re asking or the kinds of results they’re hoping to show  In that case I generally resist … I will work with it to see if there’s anything there. But often not share with them because I don’t want them to make more of it than there is there

% !TEX root =  eda-interview.tex

\section{Data Wrangling}

We observed that analysts often coupled data wrangling with exploration.
As analysts received new data, they might want to explore the data.
However, data often came from many sources
or had an improper format and size for analysis tools.
Thus, analysts had to transform the data prior to exploration.
Once they explored the data, they might discover that
they needed to further handle erroneous values or rescale the data.
Due to this coupling, some analysts even associated exploratory
analysis with data cleaning.

%\q{I guess until this point I've been associating exploratory data analysis with cleaning data. I think it's because I do both at the same time. I don't know when to clean it until I've looked at it a little bit.}

% Prior work discusses a number of challenges for data wrangling and cleaning~\cite{elmagarmid:dedup,kandel:directions,kandel:enterprise,rahm:datacleaning}.
Akin to prior work~\cite{kandel:enterprise}, several analysts % (\toverify{6}/18)
reported that they often spent the majority of their analysis time
to wrangle and clean data.
As exploration tools often lack support for some wrangling tasks,
they had to switch between tools throughout the analysis and
migrate data between these tools.
We now identify commonly observed wrangling tasks that coupled with
or impeded exploration.

\subsection{Combining Multiple Datasets}

Many analysts (12/18) had to join multiple datasets or integrate similar datasets
from multiple sources, both of which presented many challenges.
To understand the similarities and differences between datasets,
they might have to profile the datasets while combining them.
P6 and P10 complained that they often had to join data from over 20 tables.
Three analysts also had to use many scripting languages to fetch the data
from multiple different platforms.

One common challenge was the inconsistency between data sources.
P5, an astronomer, reported that different telescopes published data using
various time systems, so she spent a few days just to get
the data on the same time systems before she could combine them.
P11 also described joining data with different
levels of granularity: \inlineq{Voting data is collected at precinct level
while health data is at a state level, and population data is
served at zip code level}.

% For instance, voting data is collected at precinct level, health data is at a state level where as population difficult as she had to aggregate each dataset to be at the same granularity level first.
%P11: That scenario of the voting is challenging because voting data is done in precinct level whereas health data is at a state level, and population data, like with the census, is either served at census tract level, or at zip code, or at any other number of layers.

%\q{Random astronomy specific problem all the different telescopes use different time systems. They all record time the same way but they put it in different unit so it took me two days just to get all the different datasets in the same time system.}

\subsection{Dealing with Data Size}

% One of them considered it the biggest issue in his work.
Most analysts (15/18) had to deal with data size,
which increased data processing time, impeded sharing, or even crashed their analysis tools.
P14 mentioned that it took her a few days just to retrieve the data.
P3 noted that it was \emph{``extremely difficult to share a 250GB file''}.
% \TODO{acknowledge immaturity of tools}
Several analysts complained that large datasets did not work in R.
P11 was annoyed that his data crashed both Excel and Tableau.

The analysts applied a few strategies to handle large datasets.
Some (8/18) reduced data size by \emph{sampling} the data.
P9 and P10 noted that their challenges for sampling included
\inlineq{figuring out how large of a sample size we needed and balancing how long it would take to run} as well as
\inlineq{determining how to get meaningful and represenative samples}.

% P9 noted that deciding how to sample the data
% was also challenging as he needed to determine a sample size and figure out
% how to stratify the data so that the analysis was reproducible.

Some analysts (\verified{8/18}) also reduced data size by \emph{filtering} interesting or relevant
subsets based on their domain knowledge or suggestions from domain experts.
P15 also applied signal processing techniques to detect signals of interest
from audio data, so she could explore just the relevant data.
However, analysts might not know in advance
how to filter the data until they explored the data.

% P17 -- often, I don't know how to filter the data upfront. I’ve got all the rows and then as I go into it, either the requirements of the project change and the team helps me .. We’re going to work with a subset of the data … or I myself discover that there’s some fraction of the data that makes no sense to work with.

% \ham{Filtering has so many similarity with variable selection as the latter is simply column filtering. Not sure this is worth mentioning in some way?}

% \tabemph{Strategy: Sampling}.

% \q{We have to always decide how to sample the data in a meaningful way -- like figuring out how large of a sample size we need and how to stratify the data.}

% \tabemph{Strategy: Aggregation}.
Some interviewees (4/18) handled large datasets by \emph{aggregating} them.
One difficulty for aggregation was deciding the level of detail.
For example, aggregating time series by milliseconds could make the
aggregated data too large, while aggregating by year might eliminate
important details for the analysis. However, as analysts sometimes lacked
specific questions during exploration, they might not initially
know the right aggregation level and thus
had to re-aggregate the data many times during the exploration.

%\TODO{Challenge: Limited by community's convention}
% Ham: let's not include this because this is not purely exploration

% P2 (notes): Certain types of standard plots in some fields (e.g., Manhattan plot in genomic research) take a lot of time to generate -- Thus there is a desire to speed up with better visual encoding or with filter.
% P2 (on manhattan plot): I've come up with a way that kind of works sometimes. But again, if you're doing maybe a scatter plot isn't the best way, but it's what the field expects and does. That's the main difficulty of the genetic data, it's just how do you work with so many prints at once? How do you plot them efficiently?
% TODO: Find quote: “Scatter plot may not be best, but that’s what the field expect”

\subsection{Converting Data Formats}
\label{sec:format}

Most analysts (15/18) had to convert data into formats expected by their
analysis tools. Common formatting tasks included converting file formats
and character encodings as well as manipulating data layout such as splitting
data columns and reshaping datasets into long formats.
A common complaint was that data formatting was time-consuming.
Several analysts also complained that they had to manually
format spreadsheets that did not have rectangular shapes.

% P13 - I think we deal with different data. We don't have data sets that come from one single [type of format].

\subsection{Deriving New Forms of Data}

Many analysts (13/18) derived new forms of data
more appropriate for their analyses. Many often rescaled data by
normalizing data into certain ranges (\eg 0 to 1) or applying
logarithmic transformation to make them more normally distributed.
Several applied low-pass filters or calculated moving averages
to reduce noise in the data. P14 and P17 coded
new high-level categories from the original low-level categories.
As we will discuss in \secref{sec:unstructured}, analysts also often
derived tabular forms
of unstructured data (\eg by calculating statistics)
so they could explore and analyze the new data.

\subsection{Handling Erroneous Values}

Most analysts (13/18) had to handle data errors such as
missing data and extreme values. Handling erroneous values was challenging
since any decision to filter or impute the data
required domain knowledge and might affect downstream analysis.
Thus, analysts often explored other aspects of the data and
consulted data owners before picking a filter condition or imputation method.
Since they might later find that some errors were irrelevant to their analyses, analysts sometimes ``piled up'' errors
and kept exploring until they knew which errors were important to handle.

% \q{I actually need only a tiny fraction of the dataset. In my tiny fraction, the data is actually quite clean. All those things I thought I needed to clean up, I don't need to clean those up anymore.}

% \TODO{Analysts typically handled these erroneous values by filtering them.
% In addition, analysts sometimes imputed missing data, either by manually editing
% the data or by replacing them with statistics such as moving average.}

%\TODO{Mistake from joining / format conversion}

%P7:  when you do have data that maybe you have a number of different categories, or you have a lot of missing values, or you have strange outliers, maybe you don't even have a lot of them, but you have a few of them, I think one of the challenges is to figure out ... to even determine what you want to do with them.

% !TEX root =  eda-interview.tex

\section{Data Exploration}

Once analysts wrangled their data to have a proper format and size,
they would explore the data, which sometimes led them to acquire or wrangle more data.
%We observed that all participants regularly analyzed tabular data, while several
%participants sometimes analyzed other data types including networks, temporal sequences,
%text, audio, and images.
In this section, we first summarize observed exploration process with a focus on
tabular data, the common data form for all interviewees. We then discuss
exploration challenges including choosing variables, handling repetitive
tasks, exploring unstructured data, and determining when to stop exploration.
% how to choose variables to explore, how to explore unstructured data, as well as how to determine the end of exploration.
% Finally, we discuss the analysts' wishes for more efficient exploration process and tools.

%As tabular data is a common data form for all analysts, we first report how analysts
%explore tabular data, including how they choose variables to explore and what
%tasks they typically perform during exploration.
%%by describing typical exploration tasks and reporting how analysts explore large number of variables.
%Since several participants sometimes analyzed other data types including networks, temporal sequences, text, audio, and images, we also discuss a common challenge for exploring non-tabular data.
%Finally, we report how analysts determine the end of exploration and discuss
%\TODO{their attempts and wishes for more efficient exploration process and tools}.

\subsection{Observed Exploration Tasks}
\label{sec:ex-tasks}

Analysts usually began exploring by checking what the data
contained. For tabular data, analysts would look at table headers and,
if available, read the data's documentation.
After knowing what the data were about, they would choose aspects in the
data to explore (or stop exploring if the data were irrelevant).
As we will discuss in \secref{sec:choose}, the analysts may reduce the
number of variables if necessary.

Analysts applied various methods to examine tabular data.
To profile the data, more than half of analysts (12/18) directly
looked at the data values (\eg via a print command or spreadsheet software).
Many (8/18) computed summary statistics such as the range
and central tendencies for continuous variables and
value counts for categorical variables.
Most analysts (15/18) examined univariate distributions
with histograms and count plots.
P2 and P7 reported using box plots, while P12 used kernel density plots.
Analysts sometimes wrangled a variable during exploration, \eg by
filtering irrelevant and missing values or rescaling the variable.
% with a logarithmic transformation.

Analysts examined multivariate distributions for both profiling
and discovery goals.
They often checked certain distributions to verify their assumptions
and investigated why some assumptions did not hold true.
If their exploration goal included discovery, they would also explore
various combinations of variables to see if they could learn
interesting insights.  Some of these insights might inspire
them to further explore other relevant aspects of the data.

All analysts employed bivariate plots including bar, line,
and scatter plots. A few (3/18) used 2D histograms, frequency tables, and
contour plots. Many (11/18) also explored plots with more than two variables.
In many cases, they encoded the third variable in a plot with colors.
P4 and P16 also displayed surfaces of functions with two input variables
using 3D plots. However, P16 noted it was sometimes difficult to see relationships
from a 3D plot. To examine multiple variables at the same time, several used
scatterplot matrices. P2 and P8 also used parallel coordinate plots.
If there were too many variables, some analysts grouped variables into small batches
to avoid making the scatterplot matrices too large.
A few also grouped redundant variables identified via correlation plots.

%\q{I tend to break the data into chunks that are meaningful and also small enough that I can interpret without making the scatterplot matrix too big.}

% "chose batches of 8-12 potentially relevant variables to inspect with scatterplot matrices (SPLOMs) as it was too difficult to inspect too many variables all at once."

The analysts reported that a straightforward exploration may take a few hours to a few days. %, depending on the number of variables.
However, the data were often dirty or incomplete, requiring them to acquire or wrangle more data before they continued exploring.
Moreover, analysts often had to consult and get feedback from clients or colleagues.
However, these stakeholders might not be immediately available to help,
so the analysts had to switch to other projects while waiting.
For these reasons, exploration may take several days or even weeks.

%\LATER{domain expert initially overlooked something "obvious" to them.}

% P17: “Hey. This data you gave me, you said it’s keyed by these columns, but it seems like I also need these others.” They’re like, “Oh yeah, you're right. We forgot about that.” There’s an iterative process there with the domain experts about what their data even contains, what they think it contains and what data it needs.

%\q{If you found a data cleanliness issue, you would talk to the team, the engineering team that owned that and get whatever you found baked into the data pipeline I guess.}

%\q{If the data is clean, EDA takes less time. When I work with a dataset that's beautifully clean, somebody has already gone and cleaned it up, EDA could be hours. If I work with a dataset that's super messy and I have to use it, EDA can be days, week. It really depends.}

\subsection{How to Choose Variables to Explore?}
\label{sec:choose}

%\subsection{Exploring Large Number of Variables}
%\TODO{How about filtering?}

One common challenge was choosing variables to explore. The interviewees
generally reported that they were comfortable exploring datasets
with up to one or a few dozen variables. However, many (12/18) had to analyze
datasets with several dozens to hundreds of variables and
mentioned that the number of variable combinations to explore was a challenge for them.
% Several (7/18) mentioned
% that the number of variable combinations to explore was a challenge for them.
% due to the combinatorial explosion of variable combinations.
P16 complained that picking variables was \inlineq{too time-consuming}.
P2 said that \inlineq{choosing variables was harder than plotting itself.}
P10 even said he \inlineq{sometimes skipped plotting if there were too many variables.}

% \q{``Thinking about which variables to plot is harder than plotting itself.'' - P2}
% \q{Sometimes you try a few combinations but again, if you have many, many variables, even picking a few combinations of fields to visualize  might be too time-consuming.}

% P10: I mean at that point, if it's not possible to actually ... If you've got more than a couple of dozen (columns), then you just can't really look at all the plots efficiently. So, there I would usually just deprecate the plotting.

% \subsubsection{Variable Selection}

When there were fewer variables, analysts typically examined univariate
distributions of all variables and, if possible, all bivariate distributions.
If there were too many combinations, they often tried to
choose around 10-20 variables using a number of criteria. In addition, they
sometimes applied dimensionality reduction techniques.

All interviewees regularly applied domain knowledge to choose variables.
For profiling, they often examined variables related to their assumptions
based on prior knowledge or suggestions from involved stakeholders.
For question answering and modeling, analysts might explore variables they
considered relevant to their questions or likely to affect the dependent variables.
% To select features for modeling, some analysts explored variables that
% \inlineq{seemed more likely to affect the dependent variables} to them.
For open-ended exploration, analysts might wander through data based on what they
found interesting. Though a common difficulty was deciding what would be interesting
for the audience, several analysts noted that they often explored
relationships that might have implications for decision-making.
P11 also \emph{``drew diagrams between variables with potential relationships''}
to pick variables.

% % P11: If you don't tell me to look for something, if you say, "Just go look and see what you find," then I will give you everything that I can. I'll give you all of the different variations of what I find, but if you tell me to look for something, that's what I'm going to look for.

% https://docs.google.com/document/d/1vgYEHnMHaUCPJxdtoJrWOO3_yAt9_PVyPDtmlitcn6g/edit#bookmark=id.1eeqr7i010xp

%\q{we have to like come with the good questions on our own. Especially someone like me I’m relatively new and let’s say low in the hierarchy data scientist, so I’m not talking to like the CTO about what’s good but other people in my group have historically done that.
%
%It helps to come up with research questions. If you have somebody who is like a PM or like a person who you can say a stack level like who is motivated to find and who is motivated to ask you for the question, so then that helps like kind of guide you along. Yeah, sometimes it’s hard when you are kind of like working on your own to know if you are making something that people are going to be interested in}

% \q{Some kind of knowledge or assumption come into play, you make some assumptions that these prompters are more important, or pick the one that seems to have more engineering term or explain more of the variance'}

% Count
% Drop variables: P2, P4, P5, P6, P7, P9, P12
% Group redundant variables: P8, P12, P17

More than half of the analysts (10/18) reported criteria for dropping variables.
They often discarded variables that were parts of their datasets,
but irrelevant to their analysis.
As datasets often contained duplicate or similar variables, three analysts also used
correlation plots to group redundant variables.  For each group, they then
picked a variable that was the most reliable, having no outstanding data
quality issues, and the most understandable for their audience.

%Dropping redundant variables is also important for feature engineering projects as \TODO{...}.

% \q{There's a notion of familiarity amongst the other analysts or business or people}.
% ``No matter how many variables there are, a few variables are popular and trusted by people across. If a variable is not widely used, you would also have to establish the trust saying that the data is actually populated well, has integrity, and all.}{P8
%}

% \emph{Strategy: Build simple model to select variables}
Several analysts (7/18) applied \emph{statistics and modeling techniques}
to select variables. Some built simple models, such as shallow decision trees
or random forests, to determine important features. % (4/18)
P2 and P8 examined variables that correlated with dependent variables.
However, these approaches have some limitations. P17 noted that
industrial datasets often contained duplicated variables,
which might cause some of them to appear less important in the model building approach.
% which interfered  with variable importance in the model building approach.
P2 also noted that
\emph{``sometimes there were many things that too were correlated but not important''}.

% I’ve written methods that do things like do pairwise comparisons between the columns, doing similarity comparisons. There are lots of ways to do that. The code that I’ve written uses KDTree. I do a similarity vector comparison between these, across all the instances.

% \subsubsection{Dimensionality Reduction.}
Besides selecting variable subsets, several analysts (\verified{6}/18)
utilized \emph{dimensionality reduction} techniques to explore large number of variables.
Many used principal component analysis (PCA) and plotted
the top eigenvectors. P14 also plotted data with
t-SNE~\cite{maaten:tsne,wattenberg:tsne}. % P14
However, dimensionality reduction could lead to interpretation difficulty, as P12 noted:
\inlineq{If I have a hyper-dimension that's combining 1,000 different variables,
I can't explain to my audience what it means.}

% Ham: This one is more like exploring different type of data}
% Finally, one analyst who regularly works with genomic data that have thousands
% over variables mentioned that she explored the data by \emph{deriving new metrics}
% and explore those metrics instead.

%%%%%%%%%%%%%%%%%%%%%%%%

\subsection{Handling Repetitive Tasks}
\label{sec:repetition}

% Besides the difficulty for picking variable combination,

We found that repetitive tasks % were another factor that impeded exploration for
also impeded exploration for
many analysts (7/18). % P2, P3, P8, P13, P14, P15, P17
Some said that they often had to \inlineq{reinvent the wheel},
performing similar tasks in each exploration.
P8 also wished for a better way to visualize multiple variables at the same time:

\q{I wish there were a tool that I can just browse through a gallery of each variable's plot. It would be awesome to just browse through each of the variable's distribution and outliers, then move on to the next one.}

Despite the abundance of guidelines for data analysis and visualization,
some analysts also noted that most tools did not incorporate such knowledge
or make them easily accessible.
Thus, they had to manually apply the knowledge themselves. One common challenge,
especially for programmers, was recalling how to run specific analysis commands.
Another common complaint was the lack of good defaults in tools.
P13 complained that Matplotlib often required additional customization
to make plots look good. P17 was annoyed that many plotting libraries
dropped null values by default without indicating that some values were dropped.

%\q{If you use the default just the horizontal bar plot without any kind of arguments [in matplotlib] it would not look good.}

%\q{Value counts that default in the drops nan values and so you need to pass drop in A equal false into this to keep the nan value. When would I ever want to drop the nans from my value count? For EDA, when would I ever want to do that? Never. I want to know how often it's NaN, definitely. It's just silly to me that that's the default. It's false or that it's true.}

% \q{There's some best practices on data visualization. I feel like. Classic sources like [inaudible 01:17:53] or Bill Cleveland. There are more modern sources like Tufte or the folks with tidyverse or whatever. There a lot of best practices out there on data visualization, but they're not systematized in any a particular way to be accessible to analysts. They have to be recreated or copied for people's scripts or blocks or something. The tools do very little of that and the ones that do, like Tableau, are needing a kind of a subset of analyst themes, I guess.}

To avoid repetitive tasks and ensure that they followed best practices,
some programmers %(4/18)
compiled templates for commands they often used. % P8
P17 even wrote a script to generate a Jupyter notebook that included
basic summary plots of all variables in a given dataset and ran basic checks
for data quality issues, so he could begin exploration by browsing the notebook
without rewriting analysis commands every time.
Though templates were useful for saving analysis time,
% one common challenge for templates was that each dataset often had its own subtleties,
% one challenge was that
different datasets often had their own subtleties, so analysts needed to
adjust their templates based on the data.

% There are some like kinds of boiler plate code that like everybody uses and it’s not quite worth it to write a whole library or something, because they will be little changes that people will have to make with their specific situation. It’s nice to have a code that you could just copy paste in like customers for yourself. Definitely not having to waste time like sort of reinventing the wheel is like is good when that’s available.

% P14 --  I sort of like for myself made something like templates for when I start like a new R mark down file. Like I have certain libraries I’m always loading so like I know I’m going to be reading my TSVs with using the F read command and that’s on the data table library. It’s very useful, very fast, much better than V.CSV so yeah. I think it took me like a while to build up those templates and I could still probably make better ones. Like I guess I should probably update my templates at some point to include like libraries I’m using more now.

% There is also like common coding like kind of cook book things. Like every once in a while if I haven’t made like a histogram in a while or a bar chart or something I’ll be like, “Oh shoot wait how do I live on the X axis again because it’s like I want to change the layout of the default access labels,” or something like that

% It’s nice to have a code that you could just copy paste in like customers for yourself. Definitely not having to waste time like sort of reinventing the wheel is like is good when that’s available.

%%%%%%%%%%%%%%%%%%%%%%%%

\subsection{Exploring Unstructured Data}
\label{sec:unstructured}

While all analysts (18/18) regularly worked with structured data
(\eg tables and networks), several (7/18) sometimes analyzed unstructured data
(\eg text, audio, genomic sequences, and images).
A common challenge was the lack of methods for exploring a large
collection of unstructured data.
Thus, analysts often derived new forms of data and explored
the new data instead. P13 and P16 computed word frequencies for text data.
P2 calculated missing call rates for genomic sequences.
P15 also applied signal processing techniques to extract signals
of interests from audio data. However, when it was difficult to derive
a new form of data, the analysts might have to sample the data instead.
For example, P3 profiled a large collection of image data by directly
examining a small set of samples.

% P7 did not perform exploration and
% went straight into modeling the data when she analyzed user paths on a website
% because it was unclear for her how to explore such data.

%A few analysts calculated statistics from unstructured data such as
%word frequencies for text data and missing call rate for genomic sequence
%so they can explore and analyze these statistics instead.
%Another researcher who worked with audio data similarly applied signal
%processing techniques to extract signals of interests and analyzed
%the processed data instead.
%
%
%
%As discussed earlier in Section~\ref{sec:format},

% \Ham{Theory: limited to profiling / "spot checks" -- hard to find insights that cover the whole space}

%%%%%%%%%%%%%%%%%%%%%%%%

\subsection{When Does the Exploration End?}

As exploratory analysis is open-ended by nature, a common challenge was
deciding when the exploration should end so analysts could move on to the next tasks.
When asked how they decided to end an exploration, a \verified{handful}
of interviewees (5/18) responded that they did not always have a definite answer
if an exploration was complete. From the interviews, we found that analysts
decided to end an exploration based on multiple factors including
goal satisfaction, feedback from involved stakeholders, and time constraints.

% \q{It's a mix of seeing whether I've answered the questions we've set out to answer in a satisfactory way, and then also how much time I've spent.}

%%% Achieve the exploration goal %%%

\commonvspace

\tabemph{Goal Satisfaction}. All analysts (18/18) often ended an exploration once they satisfied with their goal.
For discovery in question answering and modeling projects, they concluded when they had an intuition on
how to formulate the answers or the models. For profiling, analysts usually stopped
when they had verified all assumptions and felt they had a good sense
of the data. Analysts might move on if they thought
they had done a sufficient job, as P17 said:
% \Ham{P17 mentioned that at some point he's getting diminishing return if he explores more.}

\q{If I reached a point where I no longer saw glaring issues, I'm done.
It does not mean the data is clean. ...
% In fact, I know that it is not perfectly clean.
However, I'm not seeing any other issues in the data,
so they're small enough that I don't need to care about them.}

% Generally it’s just you know, it shrank to below some threshold, whether that’s below 1% or below 5% or whatever. At that point, it’s diminishing returns.

% The data cleaning very much hits a point of diminishing returns and you're just done and you accept the fact that it’s still messed up in places and not perfect.

%%% Confidence about comprehensiveness %%%
The analysts' confidence whether they had done a sufficient job varied
based on the exploration goals. Analysts generally felt confident about
profiling, as P6 noted that \inlineq{just looking at distributions is not that hard.}
% P18 also said
% \inlineq{we would not do the analysis unless we felt confident [about the data].}
However, they were sometimes less confident, especially when the data were large.
P3, who profiled samples of large image collections, reported that he felt
\inlineq{confident for 90\% of the time}, but sometimes worried
that he might have missed important errors in the data.
\LATER{Detecting issues is actually hard?}
For discovery, analysts generally felt less confident as the goal was
more open-ended by nature. P5 even revealed that she never knew if she had
comprehensively explored the data when exploring a dataset she had not seen before.

% \TODO{Discuss varying degree of importance}

% Domain knowlege also played a role in the analysis's confidence.
% P11 also described a case where she felt confident that she explored everything.
% However, she did not have the domain knowledge, so she thought that her analysis
% was not so insightful.

% P3: I feel like 90% of the time, I have done a good job of screening the specimen to make sure that's high-enough quality than if I'm in certain regions. But I do always worry that sometimes, maybe, I'll miss an important part of the data.  That is a challenge for us, is figuring out how to do quality checks with this big data so that we can be confident.

% Thus, a common strategy for some analysts (\toverify{}/18) stop
% when they have shareable results.
% \q{With data that's completely new where something was observed that we've never seen before. That's harder. Because you don't know what it is and you don't know what it means. And then I think you stop doing your exploratory data analysis when you've got enough material to write a paper.}

% P5: (For open exploration) I don't think we ever know that we've comprehensively explored the data.  (But for data quality check, it’s usually good.

% P11 [For 20% that clients do not give me specific goals]  I feel confident that I've explored everything, but I don't know the domain well enough to say here's your story.

\commonvspace

\tabemph{Stakeholder Feedback}.
Since determining if they had sufficiently achieved the exploration goals could
be difficult, analysts generally performed multiple rounds of exploration
where they received feedback from team members and clients in between.
% (8/18) explicitly say this, but I suspect that the actual number is higher.
They then used feedback from team members and clients to assess if they need
to further explore the data. Some analysts described that they would stop
profiling once their clients and colleagues no longer had concerns about the data.
A few also noted that early feedback from colleagues sometimes helped them
terminate a low-value project early and let them focus on more important projects.

% P11 - It's just a lot of iterations, back and forth making sure that I haven't misrepresented anything and that what they think is really there is really there.

For open-ended discovery, analysts often ended a round of exploration
when they had shareable results. One industrial analyst (P9) mentioned that he usually
stopped when he found a result \inlineq{worth sitting down and discussing.}
P5 who used a large public data (P5) for her research mentioned that she stopped
exploring when she \inlineq{discovered enough material [to analyze] for
a paper}.  By writing a paper, she then received feedback from the
research community, driving her to do further analysis. Analysts also sometimes
stopped exploring if the data had nothing interesting.

% P9: as soon as we have an actual, like, result, like about a product that we can share ... As soon as something is polished enough to actually share with other people, and say, "Okay, this is an interesting result that's worth sitting down and talking about," that's kind of the time to start documenting better and wrapping it up.

%%% Time / Resource Limitation %%%

\commonvspace

\tabemph{Time Constraint}.
Half of the analysts (9/18) cited time limits as a major factor that prompted them to stop exploring.
P16 said that
\inlineq{it is okay to explore data for a few weeks, but after that I will need to start the other parts of the work.}
P17 also noted the pressing nature of his work:
\inlineq{we are developing models, and we have to deliver. It's happened that we have some stones left unturned---sometimes we come back, sometimes we don't.}
For time-sensitive projects, analysts might skip some parts of the exploration, as
P8 said:
\q{If I have to do it fast, I would not spend most of my time in exploratory analysis. I'll do some spot checks like just checking the ranges. I would not even look at the distributions and just go right into modeling.}
% One analyst reported that, if he has to complete the analysis quickly, he would just check the ranges of variables without looking at the distributions and go right into modeling.

Since analysts often had limited time to explore large amount of data,
it was difficult to perfectly explore all aspects of the data.
Thus, they sometimes returned to exploration after moving on to modeling.
A few of them also reported that poor modeling results led them to
further explore if any data quality issues caused the problems.

% !TEX root =  eda-interview.tex

\section{Reporting and Sharing Analysis}
\label{sec:reporting}

As data analysis is iterative and collaborative, analysts had to
share their analysis results throughout the process.
We now discuss common challenges for sharing analysis results.

\subsection{Adjusting Reports to Match Analysis Audience}

Many analysts (10/18) needed to adjust their reports to
match their analysis goals and the audience's background.
P17 mentioned that his goal was to
\inlineq{produce insights for the audience with the least amount of effort for them to understand.}
P18 also noted that \inlineq{explaining complicated things in a simple way} was the hardest thing in data analysis.
% , while
% another said \inlineq{I want to make sure my research is understandable and directly address the audience's questions. % I also want to make sure that it is as concise as possible to get my point across.
% }.

% \q{The goal is to produce insight for your audience with the least amount of effort on everybody involved. You're going for low cognitive load. If I have to introduce an entirely new metric and a whole concept from data science in order to explain to them what’s going on here, that’s painful for everybody. It’s much better if I can put it in terms they understand, use the language that they already use, talk about what they understand and know and say, okay, here’s what you have in terms of that, which may not always be what I saw in the data. It may be, oh, your telemetry system is broken. It’s producing bad values for X percent of all customers.}

We observed a few strategies for simplifying analysis reports.
First, analysts typically avoided using sophisticated plots, such as box plots,
in reports for stakeholders with less data analysis expertise.
Moreover, while their explorations might have many delicate details,
they often presented only the most important findings,
such as ones that had implications for decision-making.
% \TODO{He also uses Tableau to explore, but the exploration history provenance isn’t something he would show to the manager, but he will pick the most interesting aspect and easier to understand to show. P-18}
% \q{No matter how many variables I have. I just pick the 5-10 most important variables, and present the trends for these variables.} % P8
However, a challenge was that their analysis audience had varying
degrees of expectations. Some might even expect to explore the reports themselves, requiring the analysts to create dashboards for the reports.

% some analysts had difficulty choosing the right level of details
% because their audience had varying degrees of technical expertise,
% so some of them expected more details than others.
% \TODO{Some even want to explore a bit} % p14, P18

The need to communicate with an audience also led the analysts to align their
analyses with the audience's background. When possible,
they would choose concepts that the audience were familiar with.
As discussed earlier in \secref{sec:choose}, one criterion to choose a variable
from a group of redundant ones was whether the audience would understand it.
P8 also reported that he avoided introducing a new metric
in his analysis if there was a similar but widely-used metric.

% Analysts also collaborated with their colleagues to adjust their
% presentation. For example, a consulting data scientist (P16) mentioned that
% he always worked with a project manager to ensure that his presentation would
% be understandable to the clients and directly address their analysis goals.
% \TODO{do we have examples of people consulting their mentors, etc.?}

% Akin to other analysis activities, analysts also collaborated with their
% colleagues to ensure that their presentation would be understandable
% to the analysis audience and directly address their analysis goals.

\subsection{Analysis Sharing and Provenance}

% The interviewee also reported challenges for sharing analysis source code and results.
% For source code, diversity of tools and their interoperability hinder sharing.
% One analyst reported that she rarely shared her analysis source code for a few
% reasons.  First, her collaborators often used domain-specific software while
% she preferred to use Python and Jupyter Notebook.  In addition, she found
% Jupyter notebooks difficult to used with version control systems.
% It's also really hard to collaborate on Jupiter Notebooks. That's what I found. Just Jupiter Notebooks and Version Control don't play well together. - P5

Sharing analysis history across organization was a common challenge for many analysts (8/18),
as P14 said:
\inlineq{I often felt that I'm reinventing the wheel, but it'd take me a week to find somebody who already did something similar.}
A few also reported that their companies tried to use collaboration platforms such as a wiki
to share analysis summaries. However, these attempts eventually got abandoned because
analysts did not want the extra work to write a summary, in part because they had already
presented their analysis via other forms of reports such as slides.
%Akin to~\cite{rule:notebook},
P9 also noted the tension between doing more analysis and writing more reports:
\q{Given a fixed amount of time, do I answer more questions and go as far as I can or do I go slower and write more reports? Finding the balance is a bit hard.}
% \TODO{Powerpoint limitation: Hard include all the assumption or the data cleaning caveat...}

% \q{There were many aborted efforts to try to come up with ways to say, hey, everybody work in this table. Here are issues you should be aware of in working with this data. Amazon never managed to completely solve this problem. It was a challenge.}

% % \q{We don’t have a very good way of necessarily like sharing what everybody else is doing. Sometimes I do get the impression that I like reinvented the wheel and it will take me like a week or so to find somebody who has already done something similar}

Analysts also had to revisit their own analysis history to repeat an
analysis with new data, or to help them recall prior work when they summarized
an analysis for reporting or switched from another project.
As discussed earlier in \secref{sec:tools}, some analysts utilized
computational notebooks to keep analysis history. However,
some analysts had difficulty keeping analysis history, as P12 said:

\q{I and many other analysts I know often went through an awful lot of charts and later realized there were a few that we wanted but didn't save along the way.}

% I need a repeatable process because this exploratory data analysis that I do, it’s not a one-shot thing. If I were to do all this in a spreadsheet or as a list of random commands I execute at command line, I would be lost because now I need to go back and repeat it.

% Another stated that sharing two thousand lines of experimental R codes
% Even with programming language like R, it takes time to run.
% \q{When we typically share information at Twitter is to write PowerPoint slides, that’s like not necessarily the best way of presenting data though. It’s nice to have a nice simple graphic that’s relatively self-contained and so on, but sometimes you have a lot of caveats you have to put in there.}

% !TEX root =  eda-interview.tex

\section{Design Opportunities}

% However, our interview suggested that analysts had to perform repetitive tasks
% and face the difficulty for choosing variables during exploration.
% They also often had limited time and sometimes did not know if they had
% sufficiently explored the data. Moreover, their exploration were often interrupted
% by the need to consult other stakeholders and further acquire or wrangle data.

Based on these interview results, we now identify design opportunities
for improving data exploration tools.

\subsection{Facilitate Rapid Exploration with Automation}

From the interviews, we observe many challenges in exploratory data analysis that suggest
opportunities to augment data exploration with automation and guidance.

% Due to many challenges in exploration, we argue that tools should facilitate
% more rapid and broad exploration by providing automation and guidance. \TODO{link}

% \subsection{Facilitate Rapid Exploration with Automation}

First, as analysts often need to perform repetitive tasks and have limited time
to explore data, tools should provide automation to help analysts focus on
analyzing data rather than executing routine tasks.
% and had to explore a large number of variable combinations.
% \TODO{What people bring to the table?}
While some existing tools provide sensible defaults for plotting
commands~\cite{wickham:ggplot} or help automate chart design~\cite{mackinlay:showme},
these features are not yet available in popular analysis environments
such as Python.  More importantly, analysts still have to manually create charts one-by-one.
As we observe that some analysts apply templates to automate chart generation
and wish to browse charts without manually plotting them,
tools can recommend charts for analysts to examine\paperonly{~\cite{voyager,wills:autovis}}.

% Meanwhile,
% since datasets may have different subtleties
% that required analysts judgement, tool should allow analysts to steer
% the recommendations and adapt

% while still letting the analysts
% steer and adapt their analyses based on the subtleties of the datasets.

% Meanwhile, as different
% datasets often contain subtleties that require analysts' interpretation,
% tools that provide plot recommendation should maintaining the analysts'
% flexibility to drive the recommendation based on their interpretation.

% , as P8 said:

% \q{I wish there were a tool that I can just browse through a gallery of each variable's plot. It would be awesome to just browse through each of the variable's distribution and outliers, then move on to the next one.}

% \TODO{Exploration is a high throughput exercise.}

As analysts noted in \secref{sec:repetition}, tools can incorporate analysis practices
into their recommendations. Since analysts should begin exploring
data by examining univariate summaries
of all variables~\cite{moore:practiceofstats,seltman:experimental},
tools can suggest these plots for the
analysts\paperonly{~\cite{voyager,wills:autovis}}\thesisonly{~\cite{voyager}}.
% Ham: I think this observation is true, but want to double check first.
% Since it's not that important, I'll handle this later.
% As we observed that a few analysts did not follow common analysis practices
% to examine univariate distributions of variables, a tool may guide the analysts
% to follow the practices by suggesting such plots~\cite{voyager}.
When an analyst plots an average of a variable, a tool may augment
the plot with variance to convey uncertainty, or suggest
robust statistics such as median if there are outliers skewing the average.
For large data, a tool may suggest approximate techniques such as sampling,
online aggregation~\cite{fisher:trustme,hellerstein:online},
or density based plots such as histograms and binned scatterplots~\cite{carr:scatterplot} instead of plotting all individual points.

Another key difficulty for data exploration is choosing variables to explore.
% To effectively achieve profiling and discovery goals, analysts also had to
% broadly explore different aspects of the data.  For profiling, analysts
% typically wanted to detect data quality issues, tools should apply automatic
% routines to flag variables with potential issues such as missing values or outliers.
% \TODO{group this with scalable practice?}
% For discovery, analysts should avoid fixating on certain aspects
% of the data; otherwise they might overlook important insights in the data.
We observe that analysts heavily rely on their judgment including
determining what variables are interesting and deciding if they have
sufficiently explored the data.  One potential risk is that analysts may
be biased to focus on what they or their stakeholders are interested in,
and thus overlook important insights in the data.
Tools might reduce this risk by suggesting analysts to explore other aspects
of the data and promoting serendipitous discovery.

An important question is how to recommend data for analysts to explore.
For profiling, tools may automatically detect and suggest variables with
potential issues such as missing values or outliers~\cite{kandel:profiler}.
For discovery, suggesting data is more challenging as the goal is open-ended.
While prior work ~\cite{seo:rankbyfeature,wills:autovis,vartak:seedb}
leverages statistics for suggestions, we find that analysts mostly pick
variables based on their interpretation of semantic relationships between
variables, while statistical properties are sometimes irrelevant.
% , as our participants point out.
% (\eg by suggesting variables based on correlation~\cite{seo:rankbyfeature}
% or recommending data subsets with high deviation~\cite{vartak:seedb}),
Thus, tools should at least allow analysts to steer suggestions based on their interests.
An open question is how to design an elicitation method that lets
analysts convey domain knowledge such as how the variables could influence each other.
Tools might then store and use this information to recommend relevant variables, and
possibly help refute hypothesized relationships.
As analysts in the same organizations often explore the same datasets at different
times, tools may also leverage prior analyses to learn relevancy between variables.

% We also observed the distance between the analysts and the data domain
% and data preparation process, causing the analyst to have lack domain and
% operational knowledge necessary for their analysis.  In term of operational
% knowledge, data management tools should incorporate more information to help

% Beyond relevant variables, analysts also needed operational knowledge such as
% where the data were stored, and how the data were collected and processed.

% "Nothing — not the careful logic of mathematics, not statistical models and theories, not the awesome arithmetic power of modern computers — nothing can substitute here for the flexibility of the informed human mind... Accordingly, both [analysis] approaches and techniques need to be structured so as to facilitate human involvement and intervention." -- Tukey

% Despite many benefits for augmenting exploration tools with guidance,
% one major challenge is balancing between automation and user control.
% On the other hand, the key of exploratory analysis is the analyst's
% flexibility to make analytical decisions based on what they discover in the data.
% Excessive automation and suggestion may interrupt user's exploration
% flow~\cite{flow}, or lead users to accept recommendations and short-circuit
% their own critical thinking. Thus, while tools should aid users with
% automation, they must also retain user's flexibility to drive the analysis.

% \TODO{complacency and bias}

%%%%%%%%%%%%%%%%%%%%%%%%%%%%%%%%%%%%%%%%%%%%%%%%%%%%%%%%%%%%%%%%%%%%%%

\subsection{Support Iterative and Collaborative Workflows}

One observation is the lack of support for browsing and searching
history~\cite{heer:history} in exploration tools. If analysts can efficiently
find analyses relevant to certain datasets and variables,
they can better understand the data and avoid repeating existing work.
Moreover, since an exploration on a dataset can be lengthy, tools should also
provide interfaces to annotate important findings so that analysts
can later revisit and summarize these findings for their reports.
As analysts may not know if they have comprehensively
explored the data, surfacing variable coverage~\cite{sarvghad:coverage}
may help them identify unexplored directions and perform more
comprehensive exploration.

Another key finding is the tight coupling between exploration and
other tasks, requiring analysts to switch tools and migrate data.
Exploration tools could benefit by either providing support for other tasks such
as data wrangling \cite{tableau-prep} or tightly integrating with existing
analysis ecosystems. For example, the JupyterLab data science environment~\cite{jupyterlab}
has an extension system that can integrate an exploration tool
for Jupyter Notebook users.
Moreover, tools should consider using a shared in-memory data format (\eg~\cite{arrow})
to reduce the need to migrate data due to tool switching.
Finally, as analysts often have to create reports or presentations to share
with other stakeholders, tools can provide scaffolding to help generate
reports from existing analyses and annotations of important findings.

% \TODO{Leverage existing ecosystem for scale?}
% As acquiring new data or wrangling existing data often require rerunning the analysis,
% graphical interfaces tool should also allow users to reapply their prior analysis.

% \TODO{Domain specific}
% New techniques and tools the aim to support exploration of unstructure
% data should similarly consider these issues to maximize their benefits
% that analysts can have from their inventions.

% \TODO{Scale}

% \TODO{Even with Jupyter, analysis still need to have discipline to  preserve execution order}

\section{Conclusion}
% !TEX root =  eda-interview.tex

This \paper{} presents the results of an interview on exploratory data analysis
with 18 analysts across academia and industry.
We characterize common exploration goals:
\emph{profiling} (assessing data quality) and \emph{discovery} (gaining new insights).
Though the EDA literature emphasizes discovery, we observe that discovery only reliably occurs in the context of open-ended analyses, whereas all participants engage in profiling across all of their analyses.
We also describe how analysis goals and context affect the tasks and challenges
in exploratory data analysis. We find that analysts must perform repetitive tasks,
yet they may have limited time or lack domain knowledge to explore data.
Analysts also often have to consult other stakeholders and oscillate between
exploration and other tasks, such as acquiring and wrangling additional data.
Based on these observations, we conclude with design opportunities for data
exploration tools, such as augmenting exploration with automation and guidance.

% \yang{LOL the conclusion feels super repetitive to the abstract / intro ...}

\acknowledgments{
  We thank Interactive Data Lab members and the anonymous referees for their feedback.
  This work was done when the first author was at the University of Washington.
  This work was supported by a Moore Foundation Data-Driven Discovery Investigator Award.
}

\bibliographystyle{abbrv-doi}

\bibliography{eda-interview.bib}

\end{document}